\documentclass[conference]{IEEEtran}
\IEEEoverridecommandlockouts
\IEEEpubid{\makebox[\columnwidth]{978-1-7281-0858-2/19/\$31.00~\copyright{} 2019 IEEE \hfill}\hspace{\columnsep}\makebox[\columnwidth]{ }}
\usepackage{cite}
\usepackage{graphicx}
\usepackage{amsmath,amssymb,amsfonts}
\usepackage{algorithmic}
\usepackage{textcomp}
\usepackage{xcolor}
\def\BibTeX{{\rm B\kern-.05em{\sc i\kern-.025em b}\kern-.08em
    T\kern-.1667em\lower.7ex\hbox{E}\kern-.125emX}}

\begin{document}

\title{Quantifying the Effects of Recommendation Systems}

\author{\IEEEauthorblockN{1\textsuperscript{st} Sunshine Chong}
\IEEEauthorblockA{%\textit{dept. name of organization (of Aff.)} \\
\textit{City College of San Francisco}\\
San Francisco, United States \\
schong14@mail.ccsf.edu}
\and
\IEEEauthorblockN{2\textsuperscript{nd} Andr\'es Abeliuk}
\IEEEauthorblockA{%\textit{dept. name of organization (of Aff.)} \\
\textit{Information Sciences Institute}\\
Marina del Rey, United States \\
aabeliuk@isi.edu}
}

\maketitle
\IEEEpubidadjcol

\begin{abstract}
Recommendation systems today exert a strong influence on consumer behavior and
individual perceptions of the world. %We are investigating whether recommendation systems display signs of popularity bias through their recommendations.
By using collaborative filtering (CF) methods to create recommendations, it generates a continuous feedback loop in which user behavior becomes magnified in the algorithmic system. Popular items get recommended more frequently, creating the bias that affects and alters user preferences.  In order to visualize and compare the different biases, we will analyze the effects of recommendation systems and quantify the inequalities resulting from them.
\end{abstract}


\section{Introduction}
With the increasing amount of data available, recommender systems are becoming pervasive in many online applications in order to help users navigate huge amounts of information and influence everyday decision making.
By 2020, there will be around 40 trillion gigabytes of data, which will consist of social media data, emails, and internet searches \cite{jeffdata2019}. All this data could be easily utilized to train and update recommendation systems in online platforms, such as Facebook, Instagram, and Amazon, in order to provide better recommendations for users. 

However, because these algorithms are trained on real-world data, which may contain all sorts of social biases~\cite{mehrabi2019survey}, there are concerns about the risk of artificial intelligence (AI) exacerbating and perpetuating the present biases. For example, there is increasing evidence that online media is causing political polarization~\cite{campbell2018polarized}. This polarization, if not considered, could be reinforced by news recommendation systems that are dynamically trained as new data comes into the system. Indeed, this same phenomena can create a rich-get-richer effect that reinforces the popularity of already-popular products~\cite{fleder2009blockbuster}.
The issue comes when a recommendation system recommends choices that are universally popular, thus affecting the behavior of users in the system. These users will interact more with already popular items, creating biased data that will be used as input for retraining the algorithmic recommendations. This process is called the feedback loop. Chaney, Stewart, and Engelhardt discussed what happens during the feedback loop and how it would be problematic \cite{chaney2017algorithmic}. Because of this loop, online platforms might optimize recommendations based on what is considered popular with the majority group, which causes popularity bias and homogenizes users' interests and perceptions. Popularity bias is when popular items are recommended frequently while less popular, niche products, are recommended rarely. Needs and preferences of the minority could be undermined if these issues are not considered.

\begin{figure}
    \centering
    \includegraphics[width=\linewidth, height=6cm]{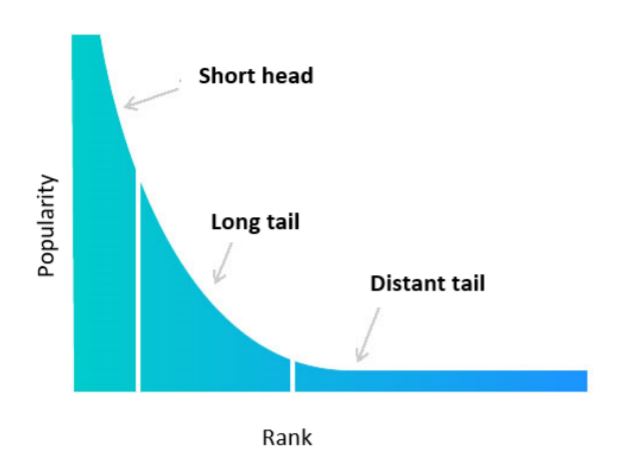}
    \caption{The short head portion contains the most popular items. Distant tail items are unpopular items that do not get recommended due to lack of positive ratings or are relatively new to the system. Long tail items have average popularity. \cite{abdollahpouri2019managing}}
    \label{fig:long-tail}
\end{figure}

Common recommendation systems, such as Amazon, are based on sales and ratings. It shows that people who bought product A also bought product B, which means novel items that have not been discovered by many other people might have a hard time surfacing and getting exposure. This tends to create the rich-get-richer effect mentioned earlier for popular items as it prevents consumers from finding better product matches because of this popularity bias \cite{fleder2009blockbuster, upenn2015}. However, increasing the discovery of novel items could also create a bias that is more personalized. For example, if we consume news through personalized recommendations, our perspective on the world would also be narrow. The key challenge is finding a balance between accuracy of the recommendation and diversity of the recommendation set. 

Figure \ref{fig:long-tail} illustrates the concept of ``long tail" items, which are items that have average popularity \cite{abdollahpouri2019managing}. These items are considered to be good, but typically do not surface in recommendations. This is because these items are not exposed to many people, so they do not have as many indicators that convey its quality to a broader audience. The \textit{x} axis represents the item rank and the \textit{y} axis represents the number of ratings per item. Indicated by the curve of the graph, ``short head" items are very popular and receive much more viewer attention.

In this paper, we will first discuss related works and their experiments (section 2). We introduce our claims that all filtering methods, besides the random method, increase inequality compared to optimal collaborative filtering (CF) recommendation systems and dynamic training has more inequality than static training (section 3). Next, we proceed to describe our methodology in conducting our own experiments to quantify the effects of inequality (section 4). To conduct our experiments we created several algorithm frameworks using CF and non-CF methods (section 5); this allows us to visualize the effects of each recommendation model and how they apply to real models used in online platforms (section 6). We will discuss how the results from our models convey inequalities in real life models (section 7) before we conclude (section 8).

\section{Related work}

Similar to our approach, researchers have explored the effects on diversity of various collaborative recommendation methods by running simulations on synthetic data and assuming different models of choice or preference for the users~\cite{fleder2009blockbuster, chaney2017algorithmic}.
Chaney, Stewart, and  Engelhardt~\cite{chaney2017algorithmic} used simulations to demonstrate how using data from users exposed to algorithmic recommendations homogenizes user behavior. They created six different recommendation algorithms with different filtering methods for the experiments: \textit{popularity, matrix factorization, random, ideal, content filtering, and social filtering.} All six of the approaches recommend from the set of items that existed in the system at the time of training. Their experiments showed that these methods did cause homogeneity through their simulated communities. A similar approach was taken in \cite{fleder2009blockbuster}, except that the choices of consumers were modeled using
the multinomial logit instead of a simple linear model of preference. 
% Their metric for measuring homogeneity was the Jaccard index. The communities each consisted of 100 users and ten new items were introduced at each time interval for 1,000 time intervals. The experiments were repeated ten times for each approach and the trials were averaged into one final result for each method. Gini coefficient was also used to determine the correlation with Jaccard index, but it was only a supplementing metric for their analysis. 
In our experiments, we do not take any assumptions on the data nor the preferences of users as we compare five filtering methods using real-world data gathered from users in \cite{jesterdata}.

% A method using a metric called Novelty Score (NS) \cite{bedi2014using} was proposed by Bedi et al. to counter popularity bias. Items are given a NS based on item frequency and inverse user frequency for unseen items. In other words, NS is only given to unseen items, which are essentially unpopular. The metric is incorporated into their proposed solution, Modified Collaborative Filtering Approach for Novel Recommendations (MCFNR). Their experiment consisted of two groups: persistent users and occasional users. Occasional users receive recommendations from the traditional collaborative filtering model while persistent users receive recommendations from the MCFNR model. The intention was to recommend relevant, less popular items to users who are using a platform often, which counters popularity bias. However, the model might only work for persistent users and not occasional users, so popularity bias would still be present to a portion of the users. 

In October 2006, Netflix released a dataset containing 100 million anonymous movie ratings and challenged the data mining, machine learning and computer science communities to develop systems that could beat the accuracy of its recommendation system, \textit{Cinematch} \cite{bennett2007netflix}. Through this challenge, Netflix became a contribution to the rise in popularity of rating prediction in recommendation algorithms. In regards to their algorithm, \textit{Cinematch} automatically analyzes the accumulated movie ratings weekly using a variant of Pearson's correlation to determine a list of ``similar" movies. After a user provides ratings, their recommendation process uses multivariate regression in real-time based on the correlations computed previously to make a personalized recommendation. The challenge primarily was concerned with accuracy rather than inequality in recommendations. The main differences between our frameworks are that we are determining ``similar" users rather than ``similar" items and that our recommendation process happens offline rather than real-time. Although accuracy is important to our algorithm, our main focus is to predict and visualize the inequalities.

Many experiments, such as a music artists study \cite{celma2008hits} and a sales diversity experiment \cite{fleder2009blockbuster}, were conducted to understand the behaviors of recommender systems and whether they affect
diversity by reinforcing the popularity of already-popular items. These empirical studies along with other previous works \cite{park2008long, shi2017long, zeng2015modeling} discuss issues of popularity bias and item inequality resulting from long-term effects of usual recommendation algorithms. As more people rely on recommendation systems as a method to discover new things, popularity bias will become more prevalent than it is currently if more online platforms use these algorithms. Research has been done to address the growing issue of popularity bias. As a result, methods were proposed to create a balance between accuracy and diversity, such as using a Novelty Score \cite{bedi2014using}, partial clustering \cite{park2008long}, and using multiple recommender systems in one framework \cite{shi2017long}. 
It is also possible to sacrifice a small fraction of short-term recommendation accuracy in exchange for higher long-term diversity \cite{zeng2015modeling}. 

The experiments in previous works were done either using more popular datasets such as MovieLens\footnote{https://grouplens.org/datasets/movielens/} and Netflix or were based on simulations. Celma and Cano's music study was based on Last.fm data, but its rating scale is ``play counts" instead of a limited range of values. In our experiment, we use Jester's dataset with \textit{Pearson correlation (Pearson), popularity (pop), matrix factorization (MF), random, and optimal} as recommendation algorithms.

\section{Problem Statement}
Real-world recommendation systems are trained using data from users who use the platforms. The problem arises when repeated recommendations could cause homogeneity because of the popularity bias and its continuous feedback loop. This is especially problematic through CF methods since one user's decisions would heavily influence those of another user. We will be focusing on analyzing and quantifying the effects of CF models to answer our claim: \textit{repeated training increases inequality in CF recommendation systems}.
\begin{figure*}
    \centering
    \advance\leftskip-1cm
    \includegraphics[width=\textwidth, height=9cm]{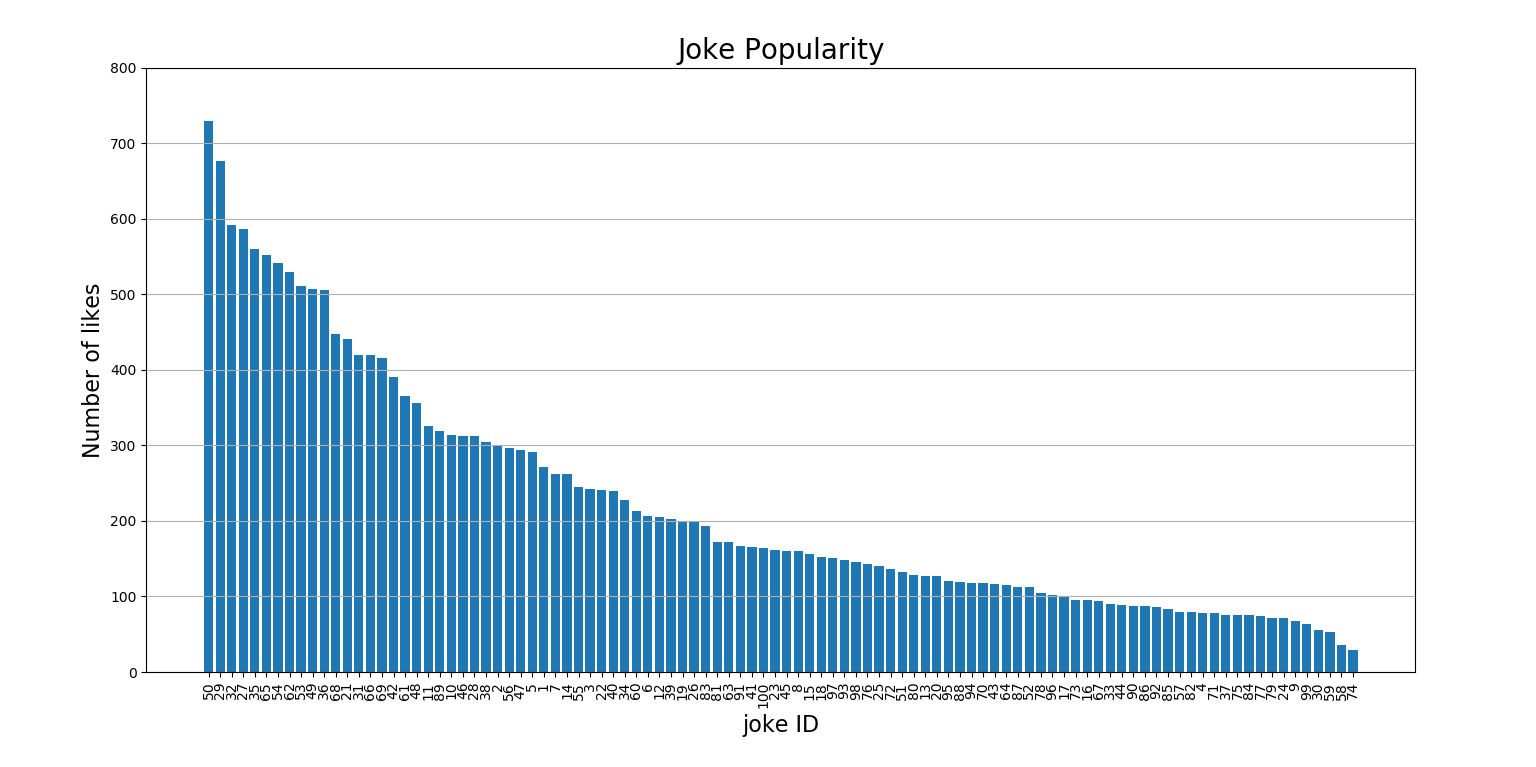}
    \caption{This graph shows the top 5 jokes that are popular for each user in Dataset 1. The shape of the graph is similar to the one in Figure \ref{fig:long-tail}.}
    \label{fig:joke_pop}
\end{figure*}

\section{Experimental Approach}
\begin{figure}
    \centering
    \includegraphics[width=\linewidth, height=5cm]{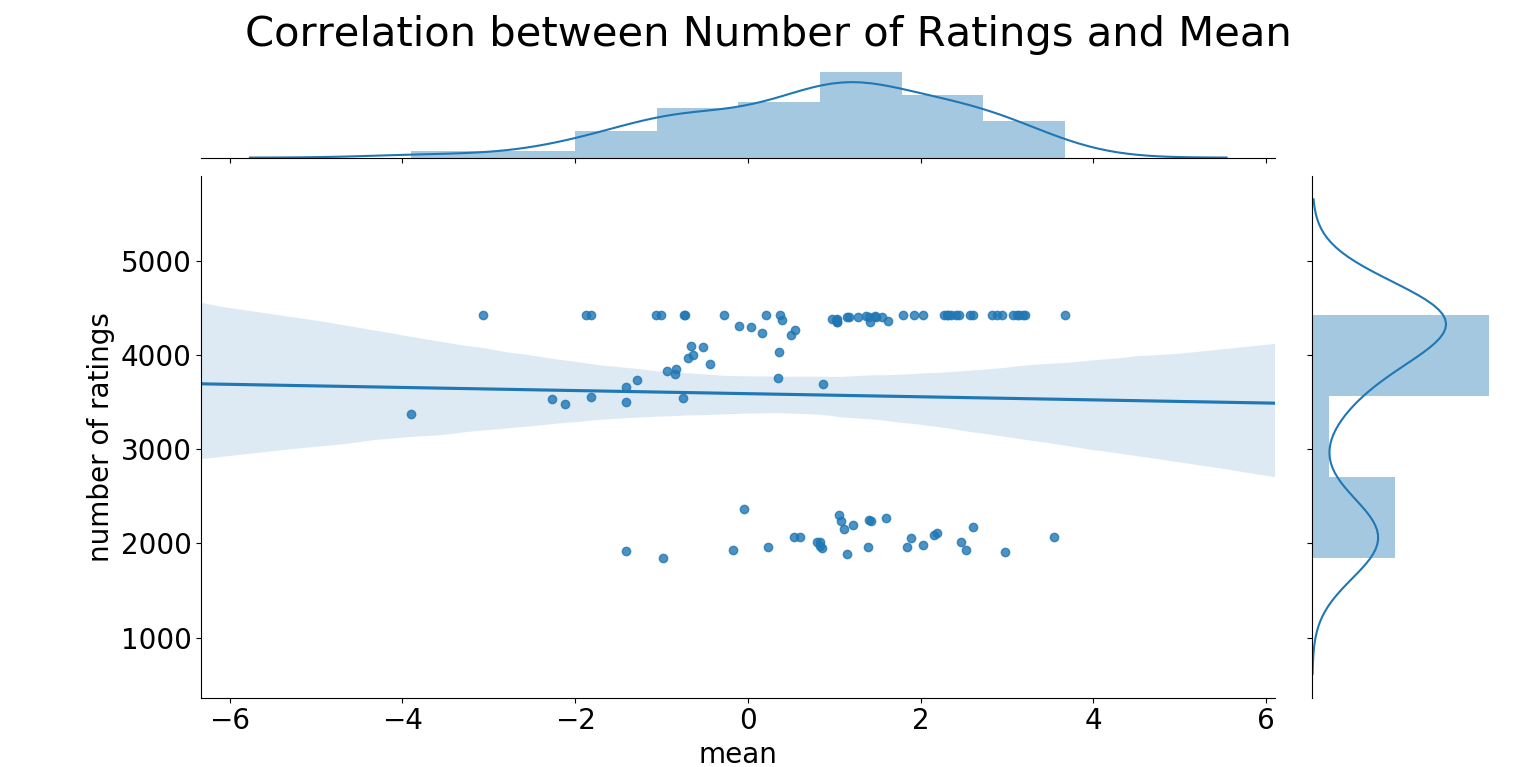}
    \caption{This graph shows that there is no correlation between the number of ratings per joke and its mean since the slope of the line is close to 0. In other words, the number of ratings does not influence the average rating for a joke. If more people rated a joke, that does not correspond to a higher mean rating.}
    \label{fig:joke_mean}
\end{figure}
We conducted offline experiments on an existing dataset from a joke recommendation system, \textit{Jester 5.0} \footnote{http://eigentaste.berkeley.edu/}, that was developed at University of California, Berkeley \cite{jesterdata}. They implemented their own CF algorithm, \textit{Eigentaste}, to make the recommendations \cite{goldberg2001eigentaste}. The dataset is a matrix that contains the following information:

\begin{enumerate}
    \item 100 joke IDs 
    \item 73,421 user IDs
    \item number of jokes a user has rated
    \item over 4.1 million continuous ratings (-10.00 to +10.00)
\end{enumerate}

We only use the first file of this dataset (Dataset 1), which contains 24,983 users and about 250,000 continuous ratings. Using the analysis on Dataset 1 will give an idea of the inequalities that might be present in their system, which could also suggest inequalities that might exist in other online platforms. A system has inequalities if the same few items are being recommended frequently, thus not all items are given equal exposure and consideration to be recommended. 

\subsection{Data}
The actual data was organized in a matrix where the rows represented the users and the columns were the jokes. There was an additional column designated to record the number of jokes a user rated. 
% In other words, Dataset 1 was contained in a 24,983 $\times$ 101 matrix. 
Users had rated at least 36 jokes. For the jokes that were unrated, their values were 99 rather than leaving them blank. The graph in Figure \ref{fig:joke_pop} shows the popularity based on the real dataset. The shape of the graph in Figure \ref{fig:joke_pop} strengthens the concept of popularity bias and ``long-tail" items mentioned in the beginning. It shows that bias is already present before conducting our experiment, which means there could have been recommendation inequality using \textit{Eigentaste}. 

Dataset 1 has no correlation between the number of ratings for a joke and its mean rating, as shown in Figure \ref{fig:joke_mean}. More ratings for a joke does not necessarily mean it has a higher mean rating. This gives an idea of how \textit{Eigentaste} made recommendations. Since it is a CF algorithm, it was focused more on recommending based on a higher mean rating rather than a high number of ratings (which could contain high and low value ratings) for a certain joke. 

\subsection{Cases}
These are the 2 cases we used for experimental analysis.

\textit{Case 1: single training}. For every test user, if the recommended joke does not have a rating, it is labeled as ``no rating," but the joke ID is still recorded. Unlike Case 2, any ratings and data on each test user are not incorporated into the training set.

\textit{Case 2: repeated training}. For every test user, if the recommended joke does not have a rating, we move to the next joke with the second best recommendation score. If we do not find a joke with a rating within the top 3 recommended jokes, then we simply do not have a rating for the test user. The new inputs that are included in the retraining of the training set are the ratings of the random jokes during the profiling phase and the rating of the recommended joke. For example, if user \textit{Y} was given 10 random jokes to rate and was given 1 recommendation afterwards, then the ratings of those 11 jokes are incorporated into the training set for retraining. We set $k = 100$, where new inputs are incorporated only after every $k$th iteration of an algorithm.

\subsection{Method}
In this section, we will lay out the general procedure for conducting our experiments. 

\textit{Reorganize data and filter users who have rated at least 50\% of the jokes.} Little data processing and cleaning was needed other than adding labels to categorize the data for easy manipulation. We used about 20,000 users who had rated at least 50\% of the jokes to decrease the chance of not having a rating for a joke that could be recommended through our system. Since we are only using the data from Dataset 1, any rating predictions we make are used as a guide and are not treated as actual data as they will not be true to the user. The analysis we are performing is reliant on true data from the user so we can preserve their true preferences that are conveyed through the previously rated jokes. Because of this aspect, we attempt to lower the chances of making a recommendation that has no rating data by taking users who have rated at least 50\% of the jokes. 

\textit{Divide users into 2 groups: training set and test set. } For the training set, we randomly picked 500 users for a smaller training set to serve as the starting point since for the filtering methods, we need data in the system first before we can make recommendations. For the test set, we randomly chose about 4000 users from the data to represent the ``new users" being added through our system. As we run this experiment multiple times, we randomly change the training and test set with different users each time. 

\textit{Create 2 different cases for filtering methods.} Case 1 is \textit{single training} (static) and Case 2 is \textit{repeated training} (dynamic). 
To quantify the effect of the feedback-loop in recommendation systems, we compare dynamic training vs a one time static training of the recommendation methods.
Both cases will start with 500 users from the training set. In \textit{single training}, the number of users in the training set stays constant during the whole experiment. Static training will only utilize ratings of the static set to recommend jokes to the test users (``new users"). In \textit{repeated training}, the starting set of users will be updated, thus the number of users in the system will not stay constant during the experiment. After every 100 recommendations, rating data from test users are incorporated into the starting set and their data will be considered in the calculation when making recommendations for the rest of the test users. Dynamic training mimics how real online platforms update data. Because of how often new users enter the system, the data that the model is making predictions on must have a similar distribution as the data on which the model was trained in order to make accurate predictions. Data distributions can be expected to drift over time, so retraining a model on newer data is common in online platforms since the data distribution could have deviated significantly from the original training data distribution \cite{aws2019}. Online platforms retrain their models periodically to ensure the accuracy of their recommendations. By using multiple methods, we are able to see the inequalities that are present and make comparisons about the impact of retraining on the system.

\textit{Run Case 1 and Case 2 each multiple times on 10 jokes for the filtering methods.} The gauge set in the data consisted of 10 jokes that every user in Dataset 1 has rated. This serves as the common set of jokes used to profile the users before making recommendations. We run the algorithm to simulate the action of giving the test users 10 jokes from the gauge set to rate. We follow the same procedure for both cases for the above methods. Based on those ratings, we compare the test user to the current database of users and its predicted ratings to generate a recommendation.

\textit{Record which jokes were recommended, the order of recommended jokes, and their ratings.} For each test user, only 1 recommendation is made. We record the joke ID and the rating of the recommended joke for each test user for both static training and dynamic training. We also record the order of the recommended jokes as it is needed to calculate the inequalities overtime. For the jokes that have no rating, we categorize them as ``no rating," but they are still part of the recommendation order and inequality calculation. For dynamic training, since we are retraining, it is ideal to obtain a recommended rating rather than no rating. More details will be discussed in the next section on how the algorithm works for the 2 cases.

\textit{Use the Gini coefficient to measure the inequalities overtime.} To measure the inequalities, we calculate the frequencies of joke \textit{i} (\textit{x\textsubscript{i}}) and joke \textit{j} (\textit{x\textsubscript{j}}) being recommended out of \textit{n} number of recommended jokes. These frequencies are then inputted into the equation defined below to calculate the Gini coefficient \textit{G}:
\begin{equation}
    G = \frac{\sum_{i=1}^{n} \sum_{j=1}^{n} \mid{x_{i} - x_{j}}\mid}{2 \sum_{i=1}^{n} \sum_{j=1}^{n} x_{j}}.
\end{equation}
The value output from the formula is between the range of 0 and 1--- 0 being \textit{perfect equality} and 1 being \textit{maximal inequality}. During the experiment, we calculate the Gini every 100 recommendations to produce a visualization of the inequality overtime. 

% \begin{figure}
%     \centering
%     \includegraphics[width=\linewidth, height=6cm]{final_static_500_2pop.png}
%     \caption{In popularity 1, the Gini over time for the popularity method stays constant overtime and is almost at maximum inequality. However for popularity 2, the popularity method shares similar Gini values to the other methods. When a system recommends the same few popular items, the inequality over time would look similar to popularity 1. When a system has many popular items (greater than 50\% of total items), then the inequality over time would look similar to popularity 2.}
%     \label{fig:2pop}
% \end{figure}

\begin{figure*}
    \centering
    \includegraphics[width=12cm, height=8.5cm]{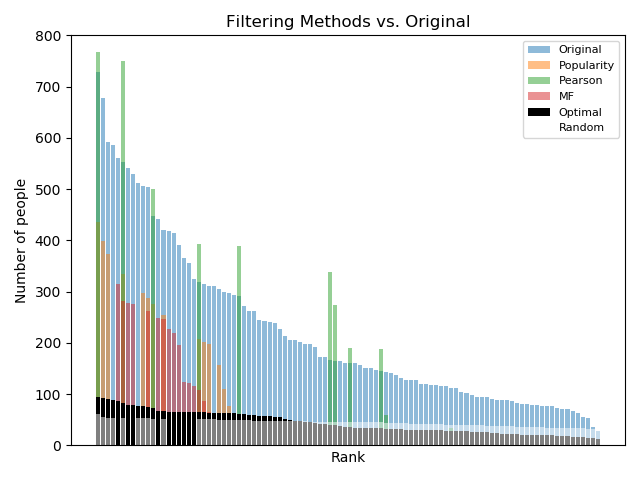}
    \caption{The original data distribution is very similar to the long-tail graph in Figure \ref{fig:long-tail}. The filtering methods, in comparison, do not have a well-defined shape as the original. The gaps represent the lack of a certain joke in the recommendation system when using a particular method, which is why some methods do not have a defined shape. The optimal method overall looks uniform compared to other filtering methods, but it is not completely uniform compared to the random method.}
    \label{fig:optimal}
\end{figure*}

\section{Recommendation Algorithms}
Next, we describe in detail all the recommendation methods analysed in our simulations.
For Pearson correlation and matrix factorization (MF), which are CF methods, the algorithms are more complex compared to the popularity, random, and optimal methods, which are non-CF methods.

\subsection{Random Method}
For this method, we expect the Gini coefficient to be a low value because the randomness of the recommendations should give each joke an equal chance at being recommended. This means that all 100 jokes should have been recommended during the experiment. The jokes are treated the same in quality, which means their ratings have no advantage when using the random method. Thus, there is very little recommendation inequality as the Gini coefficient is low. Although, there is little inequality, it is imbalanced in terms of joke quality. Chances of getting poor recommendations is high, so it is important to find a balance between recommendation inequality and joke quality. A recommendation system would not be effective if users get recommended items that do not match their preferences. Using random filtering methods increases the chances of getting poor recommendations since they are not personalized. 

\subsection{Popularity Method}
Our popularity method takes the most popular jokes, the ones with the highest average rating, and recommends one of them to the test users. In other words, a joke's probability of being recommended is proportional to its rating. This means that if a joke has a high probability of being recommended, then it has a high mean rating. We expect this type of popularity method to have the highest inequality because it only recommends a select amount of jokes that have high probabilities and high mean ratings. The Gini coefficient for this method is the highest out of all the filtering methods. However, if there are many popular jokes, such that among the 100 jokes, the majority had similar ratings that exhibit their popularity, then the inequality would not be as obvious. 
% Figure \ref{fig:2pop} shows an example during static training where popularity methods might not always have high inequality. 
In the real world, recommendation systems do not always recommend only a few popular items. Depending on the platform and the amount of user data it has, popularity could span to a majority of items, which could lower the overall recommendation inequality.

We will not be focusing on the popularity method with maximal inequality mentioned previously as most online platforms would not use this type of algorithm. Instead, we will be using these two popularity methods: 1) one that uses the joke's mean rating scaled to an exponent of 1 and the other that uses the joke's mean rating scaled to an exponent of 2. The methods are defined by the equation $P_j$ below: 
\begin{equation}
    P_j = \frac{(r_j)^k}{\sum_{i = 1}^{n} (r_i)^k}
\end{equation} 
$P_j$ is the probability of joke $j$ being recommended, $r_j$ is the mean rating of joke $j$, $k$ is the exponent, i.e 1 or 2, and $\sum_{i = 1}^{n} (r_i)^k$ is the sum of the scaled mean ratings of $n$ jokes. Scaling it to exponent 2 increases and exaggerates the probability of a joke being picked based on its popularity, making it even more likely to pick a popular joke but not restricting it to only a select amount of popular jokes. 

\subsection{Pearson Correlation}
We modified a simple CF Pearson correlation algorithm written in 2012 by Wai Yip Tung \cite{wai2012}. The main idea is that we compute the \textit{similarity} score between a user \textit{X} from the training set for all users and a user \textit{Y} from the test set. Using that \textit{similarity} score as the weight, we calculate a value, in which we label as the \textit{similar rating} score, by combining the weight with each joke rating. We repeat this with all users in the training set. In other words, each test user will have at least 500 \textit{similarity} scores, one for each user in the training set, and 100 \textit{similar rating} scores, one for each joke. Although 100 \textit{similar rating} scores are given, we only use at most 90 ratings since we exclude the gauge set. In the original dataset from \textit{Jester 5.0}, the gauge set was not part of the recommendations so we excluded it so our recommendation process would be similar to \textit{Jester}. The \textit{similarity} score is calculated using Pearson correlation \begin{math}\rho\end{math}X,Y, which is defined as:
\begin{equation}
    \rho X,Y = \frac{cov(X, Y)}{\sigma_{X}\sigma_{Y}}.
\end{equation}
Covariance \textit{cov} for \textit{n} total ratings is
\begin{equation}
    cov(X, Y) = \frac{\Sigma(X_{i} -  \bar{X})(Y_{j} -  \bar{Y})}{n}
\end{equation}
where \textit{X\textsubscript{i}} are the ratings of user \begin{math}\textit{X}\end{math}, $\bar X$ is the mean of \textit{X\textsubscript{i}}, \textit{Y\textsubscript{j}} are the ratings of user \textit{Y} and $\bar Y$ is the mean of \textit{Y\textsubscript{j}}. The covariance is the joint variability of the 2 users. Pearson correlation gives a more accurate range (from 0 to 1) on user similarity. Close to 0 means that users have different preferences and close to 1 means users have similar preferences.

To calculate and normalize the \textit{recommendation} score for each joke, we divide the sum of the \textit{similar rating} scores by the sum of the \textit{similarity} scores. We sort the jokes with the highest recommendation score and exclude any recommended jokes that are part of the gauge set. The joke with the highest recommendation score is recommended. There are some cases where the top recommended joke does not have a rating recorded in Dataset 1. This is because that our CF algorithm behaves differently and would not necessarily give the exact same recommendations as \textit{Eigentaste}. Nevertheless, our algorithm handles that aspect differently in both cases.

\begin{figure}
    \centering
    \advance\leftskip-1cm
    \includegraphics[width=11cm, height=6cm]{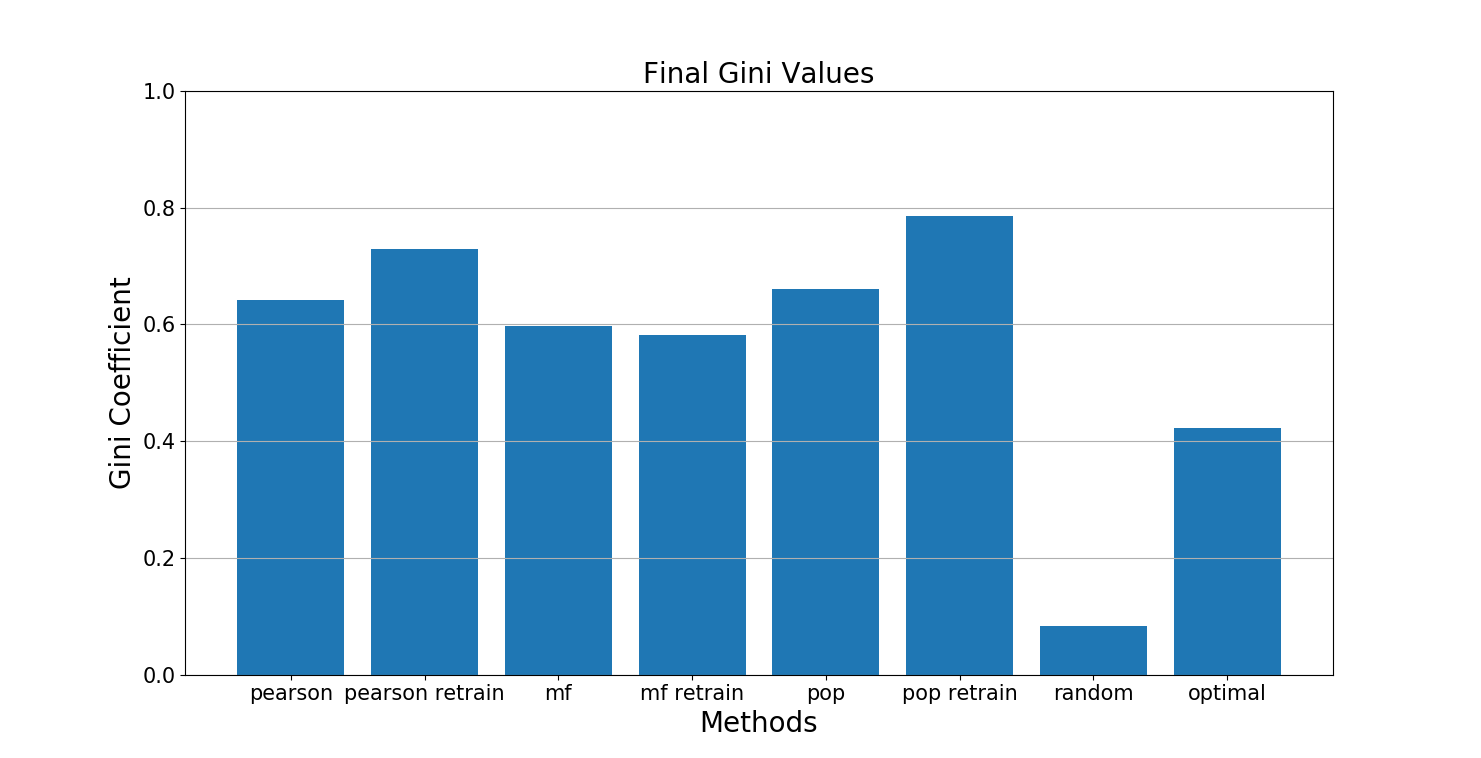}
    \caption{Comparing the Gini values for various methods, this shows that recommending based on popularity while retraining yields the highest Gini coefficient. The popularity method used in this graph is popularity 2, which is the one that uses the joke's mean rating scaled to an exponent of 2.}
    \label{fig:final_gini}
\end{figure}
\begin{figure*}[!tbp]
    \centering
    \begin{minipage}[b]{.45\linewidth}
        \includegraphics[width=\linewidth, height=6cm]{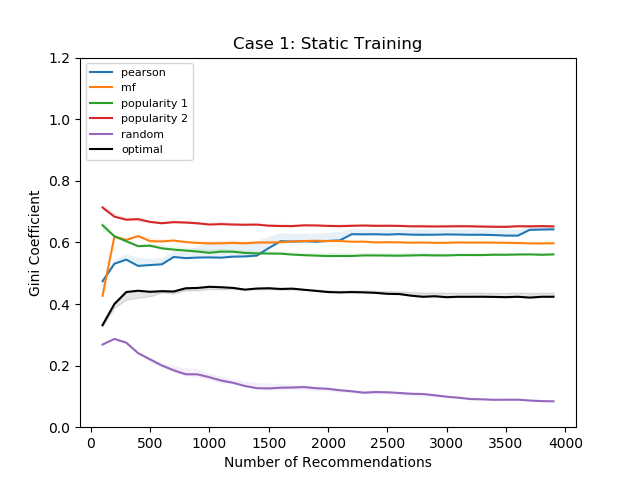}
    \end{minipage}
    \begin{minipage}[b]{.45\linewidth}
        \includegraphics[width=\linewidth, height=6cm]{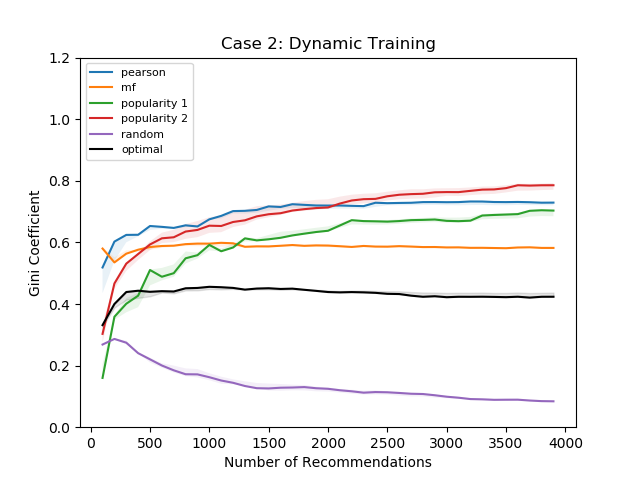}
    \end{minipage}
    \caption{In static training, the Gini values converge and stabilize fairly quickly since the slopes of the graph are close to 0. However, the Gini value overtime grows at a quicker rate during dynamic training. The increased rate of growth in dynamic training indicates its potential for reaching high inequality, especially for the popularity methods.}
    \label{fig:gini_graph}
\end{figure*}

\subsection{Matrix Factorization}
For MF, we modified another simple algorithm by Nick Becker in 2016 \cite{nick2016}. The process of using the training set and test set is similar to the CF Pearson algorithm, but the way it creates recommendations is different. This algorithm utilizes singular value decomposition (SVD), which decomposes a matrix $R$ to a smaller approximation of the original matrix $R.$ Mathematically, it decomposes 
$R$ into two unitary matrices and a diagonal matrix:
\begin{equation}
    R = U\Sigma V^T
\end{equation}
where $R$ is a user ratings matrix, $U$ is the user ``features" matrix, $\Sigma$
is the diagonal matrix of singular values (weights), and $V^T$ is the joke ``features" matrix. $U$ represents how much users ``like" each feature and $V^T$ represents how relevant each feature is to each joke. To get the lower rank approximation, we take these matrices and keep only the top $k$ features, which we think of as the 
$k$ most important underlying taste and preference vectors \cite{nick2016}. In other words, these features are not explicitly passed when computing the recommendations. It is through matrix multiplication that it had generated some hidden ``features" and picked up underlying preferences of users. After multiplying the 3 matrices, we get prediction ratings for each user. If prediction ratings for a user are high for certain jokes, then we recommend those jokes to the user.

\subsection{Optimal Method}
We use this model to compare against the real preferences of people. It serves as the baseline for what people actually like, which encourages recommendation systems to improve their frameworks to meet people's expectations. In the \textit{optimal} method, the Gini coefficient stabilizes at a value very quickly. Optimal does not necessarily mean perfect equality, but in a sense that all users like the items being recommended to them. Note that this is very difficult in the real world. Recommendation algorithms will recommend items close to a user's preferences, but they are not always accurate. There are times where users will reject the recommendations because they did not match their interests.

Since we defined the optimal case as having all users like the items being recommended, we assume those items have the highest ratings. Thus, the method is defined as:
\begin{equation}
    S_u = \{s_i\}_{i = 1}^{n}
\end{equation}
where $S_u$ is a set of ratings for each user $u$. There are a total of $n$ joke ratings for each user. To find the optimal joke $o_u$ for user $u$, we find the highest or maximum rating in $S_u$, which is defined below:
\begin{equation}
    o_u = max \ S_u
\end{equation}.

Unlike the cases for single training and repeated training, all recommended jokes will have ratings recorded previously because we are taking the highest rating that exists in Dataset 1 for each test user. For each user in this experiment, we find the joke that has the highest rating and record the joke ID and its rating. From here, the procedure for finding the Gini is the same as the above cases. 

\section{Results}
In this section, we compare the experiments that resulted from our algorithms and evaluate the implications of the results.

In Figure \ref{fig:optimal}, the optimal method is compared to all the other methods as well as the original data distribution. Comparing the graph to the earlier graph in Figure \ref{fig:joke_pop}, the shape of this graph shows it is less biased by a huge degree. It is not perfect uniformity, as that would be impossible in the real world. Making it completely uniform would be difficult and unrealistic as there would have to be a constant number of people who like each joke. By having the optimal model, it creates a baseline for comparison when analyzing the various methods.

Figure \ref{fig:gini_graph} shows similar patterns of increasing inequalities in both static and dynamic training for Pearson, MF, and popularity methods compared to the optimal method. In single training, most methods stayed relatively with the same inequality as more recommendations were made overtime. The slopes of the lines are low, which shows that they are stabilizing in inequality with their recommendations. For the optimal, it appears to be higher in inequality compared to random, but, as mentioned earlier, recommendation inequality is not the only factor in determining an effective recommendation method. If inequality is too low, then users would frequently get recommendations that might not correspond to their preferences. In dynamic training, the graph shows a clear increase in growth for most of the methods, especially the popularity methods. The growth in Pearson is not as obvious, but its Gini and slope is greater than its counterpart in static training. Since various recommendation methods operate differently and can be more robust than others, their approaches can vary in inequality. Little changes between static and dynamic training could mean that the method is more robust than others. Nevertheless, in both cases, these recommendation methods have higher inequality than the optimal during the majority of the recommending process.  

In Figure \ref{fig:final_gini}, we recorded the median Gini values for multiple trials of the experiment. The graph for the optimal Gini is around 0.4. From this experiment, we define the Gini of about 0.4 as the optimal case if a recommendation system recommends jokes that every user would like. In other words, giving recommendations that satisfies every user's unique preferences shows that a diverse set of jokes are being recommended in the optimal case. Looking at Figure \ref{fig:final_gini}, the final Gini value of the optimal is the lowest, not including the random method, which proves that our claim that various filtering methods increase inequalities in CF recommendation systems.

\section{Discussion}
Based on our results, we could apply our findings to other datasets. For example, if we had a dataset of user ratings or likes from online platforms, such as Amazon, Instagram, Netflix,  etc., analyzing their datasets will show if there is recommendation inequality present in their systems. This experiment works best with platforms that utilize CF models, but it could be modified for other types of filtering methods. Following our definition of the optimal case, the ideal recommendation system for live platforms might not have a Gini of exactly 0.4, but it should be lower compared to the Gini value of the actual recommendation framework. Our results suggest that real recommendation systems have plenty of room for improvement.

\section{Conclusion}
An exploratory analysis of a CF dataset visualizes the impact of recommendation systems on users. We found that common recommendation methods such as Pearson, MF, and popularity, have more inequality in recommendations than the optimal and dynamic training has more inequality than static training, which reinforces the idea of feedback loop (section 6). These findings create awareness for designers and consumers in live recommendation platforms. Designers need to consider how these inequalities will affect their users. If these effects are not addressed, they will be amplified and become problematic as more users conform to interacting with only certain recommended items, creating the lack of diversity. Every system will have some bias; the best we can do is to reduce it. Every dataset is represented differently, and using certain recommendation methods might increase recommendation inequality. The first step in solving this issue is to increase awareness that using different recommendation methods affects the amount of inequality present. Once people realize that, then they can make individual choices or propose solutions for these recommendation algorithms in order to achieve the balance in quality and diversity shown in the optimal method.

\section*{Acknowledgements}
This work was performed under REU Site program and funded by NSF grant \#1659886.

\bibliographystyle{plain}
% \bibliography{citations.bib}

\vspace{12pt}
\color{red}

\end{document}